\documentclass[aps,prd,preprint,superscriptaddress,amsmath,amssymb,nofootinbib]{revtex4-1}
\pdfoutput=1
\usepackage{graphicx,hyperref,xcolor,bm}
\allowdisplaybreaks

\newbox\slashbox \setbox\slashbox=\hbox{$/$}
\newbox\Slashbox \setbox\Slashbox=\hbox{\large$/$}
\def\pFMslash#1{\setbox\@tempboxa=\hbox{$#1$}
  \@tempdima=0.5\wd\slashbox \advance\@tempdima 0.5\wd\@tempboxa
  \copy\slashbox \kern-\@tempdima \box\@tempboxa}
\def\pFMSlash#1{\setbox\@tempboxa=\hbox{$#1$}
  \@tempdima=0.5\wd\Slashbox \advance\@tempdima 0.5\wd\@tempboxa
  \copy\Slashbox \kern-\@tempdima \box\@tempboxa}

\newcommand{\eps}{\epsilon}
\newcommand{\euv}{\epsilon_{\mathrm{UV}}}
\newcommand{\eir}{\epsilon_{\mathrm{IR}}}

\newcommand{\dms}{\frac{(\mu^2 e^{\gamma_{\mathrm{E}}})^{\epsilon}}{\Gamma(1-\epsilon)}}

\newcommand{\ddk}{\frac{d^D k}{(2\pi)^D}}
\mathchardef\mhyphen="2D

\newcommand{\as}{\alpha_s}
\newcommand{\asc}{\frac{\alpha_s C_F}{2\pi}}

\newcommand{\msbar}{\overline{\text{MS}}}
\newcommand{\kt}{\mathbf{k}_{\perp}^2}
\newcommand{\mud}{\tilde{\mu}^{2\eps}}

\begin{document}

\title{Anatomy of the Quasi-PDF with a Transverse-Momentum Cutoff}

\author{Junegone Chay}
\email{chay@korea.ac.kr}
\affiliation{Department of Physics, Korea University, Seoul 02841, Korea}

\begin{abstract} \baselineskip 3.5 ex 
The quasi-PDF is fundamentally different from the light-cone PDF, and large-momentum effective theory relates the
two through perturbative matching. This has required considerable effort. One of the difficulties in matching is
that no single
renormalization scheme has been
agreed upon for the quasi-PDF, and the resulting variety of prescriptions can obscure the physics underlying the
matching. We study how the choice of scheme affects the renormalized quasi-PDF and the matching coefficient,
and how the two are related, aiming to identify a scheme in which this dependence becomes transparent. As a first
attempt, we consider the transverse-momentum cutoff scheme. It is not used in practice, since a hard momentum cutoff
breaks Lorentz invariance, and it has generally been treated as an illustrative example exhibiting the linear
divergence of the spatial Wilson line. A careful study, however, reveals a richer structure than expected. The
ultraviolet divergence in this scheme depends not only on the cutoff itself, but also on an independent divergence
associated with the momentum fraction of the quasi-PDF. Disentangling these two ultraviolet structures, and
following the disentangling through the explicit one-loop calculation, uncovers further structure along the way.
Once the ultraviolet contributions are correctly separated and assigned to the renormalization counterterm, a finite
ambiguity remains, and we use it to restore the quark-number conservation broken by the cutoff. The resulting
renormalized quasi-PDF and matching coefficient can be meaningfully compared with the result obtained in
dimensional regularization, and we show how the two are reconciled.
\end{abstract}

\maketitle

\section{Introduction}
Parton distribution functions (PDFs) are essential nonperturbative elements for high-energy
 processes~\cite{Ellis:1996mzs,Collins:2011zzd}.  They enter factorization formulas involving hadrons, where the 
 short-distance partonic scattering cross section can be computed perturbatively, while the long-distance structure of the hadron is 
 encoded in PDFs. They describe the longitudinal momentum distribution of partons inside hadrons and 
are universal in the sense that the same PDFs enter a wide class of scattering processes. Their scale dependence is 
governed by perturbative evolution~\cite{Gribov:1972ri,Gribov:1972rt,Dokshitzer:1977sg,Altarelli:1977zs}, but the PDFs 
themselves are basically nonperturbative objects.

This makes it important to obtain information on PDFs from first principles.  Lattice QCD is the natural framework for 
nonperturbative QCD, but the definition of the PDF involves a lightcone separation of quark fields.  Such a lightcone 
correlation is not directly accessible in a Euclidean lattice calculation. One has to relate the lightcone matrix element 
defining the PDF to another hadronic matrix element in Euclidean spacetime, whose connection to the PDF should be 
theoretically established. Large-momentum effective theory (LaMET)~\cite{Ji:2013dva,Ji:2014gla} provides a way around 
this difficulty by relating lightcone PDFs to quasi-PDFs, which are defined by equal-time spatial correlations in a hadron 
boosted with large momentum $P_z$. For reviews on LaMET, see Refs.~\cite{Ji:2020ect,Cichy:2018mum}. For 
$P_z\gg \Lambda_{\text{QCD}}$, the quasi-PDF $\tilde q$ can be matched perturbatively onto the lightcone PDF $q$, up to power 
corrections~\cite{Ji:2013dva,Ji:2014gla,Xiong:2013bka,Izubuchi:2018srq}. The factorization relation can be written as
\begin{equation}
\tilde q(x,P^z,\mu) = \int_{-1}^{1}\frac{dy}{|y|} \,C\left(\frac{x}{y},\frac{\mu}{|y|P^z}\right) q(y,\mu)
+\mathcal O\left(\frac{\Lambda_{\rm QCD}^2}{(P^z)^2}\right).
\label{eq:intro-factorization}
\end{equation}
Here $\mu$ is the renormalization scale, and the matching coefficient $C$ is perturbatively calculable. The quasi-PDF and the PDF 
have the same collinear infrared (IR) divergences~\cite{Xiong:2013bka,Izubuchi:2018srq}. These divergences cancel in the matching, 
so the matching coefficient $C$ is infrared finite. This cancellation, however, does not by itself fix the ultraviolet (UV) behavior of the
quasi-PDF. The quasi-PDF is defined by a spatial nonlocal operator, and its renormalization prescription determines
which UV contributions are assigned to the counterterm before the matching to the lightcone PDF is performed. Both the
renormalized quasi-PDF and the finite matching coefficient therefore depend on this choice of scheme, and the two are
not independent of each other. The scheme dependence carried by the renormalized quasi-PDF is compensated by the
matching coefficient, and the same interplay appears in the renormalization-group equations relating them.

For the PDF, the $\msbar$ scheme has become a standard renormalization prescription. The simplest idea for the
quasi-PDF is to adopt the same scheme, applying $\msbar$ on both sides of the matching. This idea is straightforward,
though the resulting one-loop matching calculation turns out to be nontrivial~\cite{Izubuchi:2018srq}. In the
$\msbar$ scheme, dimensional regularization sets power divergences to zero, so the renormalized quasi-PDF does not
exhibit the linear divergence that the spatial Wilson line has from its own self-energy. This divergence does not
appear in the matching between the quasi-PDF and the PDF. It has not disappeared, however. It must still be confronted
in the second matching, between the quasi-PDF and the spatial correlation function computed on the lattice, where the
corresponding divergence remains.

A number of other renormalization schemes for the quasi-PDF have instead been developed from the perspective of
lattice QCD~\cite{Martinelli:1994ty,Constantinou:2017sej,Alexandrou:2017huk}. These include the
regularization-independent momentum-subtraction (RI/MOM) scheme~\cite{Stewart:2017tvs}, the ratio
scheme~\cite{Orginos:2017kos,Radyushkin:2017cyf}, the hybrid scheme~\cite{Ji:2020brr}, and the pseudo-PDF
approach~\cite{Radyushkin:2017lvu,Radyushkin:2017ffo}. These schemes are designed to renormalize the spatial
correlation function computed on the lattice, where the linear divergence of the Wilson line is genuinely present,
unlike in dimensional regularization. The divergence is removed there directly, through a Wilson-line mass
renormalization~\cite{Chen:2016fxx,Green:2017xeu} or an equivalent nonperturbative subtraction, so the resulting renormalized 
quasi-PDF, like its $\msbar$ counterpart, no longer carries a linear divergence by the time it is matched to the PDF.

A different scheme, the transverse-momentum cutoff (TMC), is not used in practice, since a finite cutoff explicitly breaks Lorentz
invariance, and the one-loop TMC calculation of Ref.~\cite{Xiong:2013bka} has generally been treated as an illustrative
example rather than a candidate for lattice renormalization. Unlike any of the schemes above, however, the TMC scheme
leaves the linear divergence of the Wilson line exposed in the continuum quasi-PDF. A hard momentum cutoff, unlike
dimensional regularization, does not set power divergences to zero, and no separate nonperturbative subtraction is
performed to remove it beforehand. The linear divergence must therefore be confronted directly in the matching to the PDF, through a procedure analogous
to the Wilson-line mass renormalization used in lattice QCD, before the usual matching onto the logarithmic structure
of the PDF can proceed. This makes the TMC scheme a useful, if unconventional, setting for studying how such
scheme-dependent structure can be separated from the physics of the matching.

A transverse-momentum cutoff regulates the ultraviolet behavior of the loop momentum. However, we have to note that
the quasi-PDF has a peculiar characteristic, that the momentum fraction $x$ can extend beyond the physical region $0<x<1$. 
Because $x$ can extend to infinity, it is possible that a divergence exists there, one that the cutoff $\Lambda$ has no power to
control. If such a divergence is present, it constitutes a class of ultraviolet behavior that the transverse-momentum
cutoff, by itself, cannot handle. It is therefore important to identify the nature of this possible divergence, and to
find a way to disentangle it from the divergence already regulated by $\Lambda$. We find that such a divergence does
exist, and that even in this simplest of cutoff schemes, the ultraviolet behavior of the quasi-PDF separates into two
independent structures.

Separating these two structures still leaves a piece of the counterterm undetermined. It is important because
introducing the cutoff breaks the Ward identity associated with quark-number conservation, and nothing in the cutoff
scheme itself requires this to be repaired. Yet standard field-theory technique tells us that a Ward identity broken
by regularization can be restored in the renormalized theory, through an appropriate choice of the finite
counterterm~\cite{Bonini:1993kt,Grassi:1999tp,Bochicchio:1985xa}. We use exactly this freedom, and choose the finite counterterm 
so that quark-number conservation is restored. We also examine how the different ultraviolet contributions identified in this way enter 
the renormalization-group equations for the quasi-PDF and the matching coefficient.
 
The paper is organized as follows. In Section~\ref{qpdf-formulation}, we formulate the lightcone PDF and the quasi-PDF and 
establish the conventions used in this paper. In Section~\ref{sepdep}, we examine the ultraviolet structure of the cutoff-regulated 
quasi-PDF and develop a general method for separating the divergence controlled by the transverse cutoff from the divergence 
associated with the large-$|x|$ behavior of the momentum-fraction distribution. In Section~\ref{matchone}, we 
apply this prescription to the one-loop quasi-PDF, construct the counterterm and the renormalized quasi-PDF, and match the 
latter to the $\msbar$ lightcone PDF. In Section~\ref{rg-tmc}, we discuss the renormalization-group behavior of the 
renormalized quasi-PDF and the matching coefficient. We conclude in Section~\ref{conc}. In 
Appendix~\ref{app:full-line-distribution}, we collect the definitions of the distributions used in the paper and derive the 
endpoint and infinity contributions needed to define the quasi-PDF as a distribution on the full line.
  
\section{Formulation of the quasi-PDF}
\label{qpdf-formulation}

In this section, we collect the operator definitions, renormalization relations, and factorization formulas used in our calculation.   
For a hadron with momentum $P^\mu$, the bare quark PDF is defined by the lightcone correlation~\cite{Collins:2011zzd}
\begin{equation}
q_{\text{bare}}(x) = \int\frac{d\xi^-}{4\pi}\, e^{-ixP^+\xi^-} \langle P| \overline{\psi}(\xi^-)\gamma^+ W(\xi^-,0)\psi(0)
|P\rangle_{\rm bare},
\label{eq:pdf-bare-def}
\end{equation}
where
\begin{equation}
W(\xi^-,0) = {\cal P}\exp\left[ -ig\int_0^{\xi^-}d\eta^-\,A^+(\eta^-) \right]
\label{eq:pdf-wilson}
\end{equation}
is the lightlike Wilson line that renders the bilocal operator gauge invariant.  We use the convention in which $-1<x<0$ describes the 
antiquark contribution.

The quasi-PDF is defined from the corresponding equal-time spatial correlation~\cite{Ji:2013dva,Ji:2014gla}.  For $P^\mu=(P^0,0,0,P^z)$, we write
\begin{equation}
\tilde Q_{\text{bare}}(z,P^z) = \frac{1}{2P^z} \langle P| \overline{\psi}(0,0_\perp,z)\gamma^z W_z(z,0)\psi(0) 
|P\rangle_{\text{bare}},
\label{eq:quasi-correlator-bare}
\end{equation}
with the straight spatial Wilson line
\begin{equation}
W_z(z,0) = {\cal P}\exp\left[ -ig\int_0^z dz'\,A^z(z') \right].
\label{eq:quasi-wilson}
\end{equation}
The bare quasi-PDF is its Fourier transform,
\begin{equation}
\tilde q_{\text{bare}}(x,P^z) = P^z\int_{-\infty}^{\infty}\frac{dz}{2\pi}\, e^{ixP^zz}\, \tilde Q_{\text{bare}}(z,P^z)
= \int_{-\infty}^{\infty}\frac{dz}{4\pi}\, e^{ixP^zz} \langle P| \overline{\psi}(0,0_\perp,z)\gamma^z W_z(z,0)\psi(0)
|P\rangle_{\text{bare}}.
\label{eq:quasi-bare-def}
\end{equation}
At finite $P^z$, $\tilde q_{\text{bare}}$ has support on the entire real line.

We renormalize the lightcone PDF in the $\msbar$ scheme.  In momentum-fraction space the bare and renormalized PDFs are related by
\begin{equation}
q_{\rm bare}(x,\epsilon) = \int_{-1}^{1}\frac{dy}{|y|}\, Z_q^{\msbar}\left(\frac{x}{y},\epsilon,\mu\right) q_{\msbar}(y,\mu).
\label{eq:pdf-ren-general}
\end{equation}
For the quasi-PDF it is convenient to state the renormalization first in coordinate space.  The straight Wilson-line operator is 
multiplicatively renormalizable~\cite{Chen:2016fxx,Ishikawa:2017faj,Ji:2017oey,Li:2018tpe}, and in a scheme $S$ we write
\begin{equation}
\tilde Q_{\text{bare}}(z,P^z;{\text{reg}}) = Z_{\tilde Q}^{S}(z;{\text{reg}},\mu)\, \tilde Q_{\text{ren}}^{S}(z,P^z,\mu).
\label{eq:quasi-ren-general}
\end{equation}
Here $\text{reg}$ denotes generically the regularization parameters of the scheme $S$ --- a single parameter in some
schemes, such as $\epsilon$ in $\ msbar$, but more than one in others, since the TMC scheme discussed in Sec.~\ref{sepdep} requires 
both the transverse-momentum cutoff $\Lambda$ and the dimensional regulator $\epsilon$ simultaneously. The corresponding 
relation for the quasi-PDF follows by Fourier transformation.  In the calculation below the transverse-momentum cutoff regulates the 
large-$k_\perp$ region of the quasi-PDF, while dimensional regularization is used for the collinear IR singularity.  The lightcone PDF is evaluated in dimensional regularization and renormalized in the $\msbar$ scheme.

For large $P^z$, the renormalized quasi-PDF satisfies the LaMET factorization formula~\cite{Ji:2013dva,Ji:2014gla,Izubuchi:2018srq}
\begin{equation}
\tilde q_{\text{ren}}^{S}(x,P^z,\mu) = \int_{-1}^{1}\frac{dy}{|y|}\, C_S\left( \frac{x}{y}, \frac{\mu}{|y|P^z} \right)
q_{\msbar}(y,\mu) + {\cal O}\left( \frac{\Lambda_{\text{QCD}}^2}{(P^z)^2} \right).
\label{eq:matching-general}
\end{equation}
This is the same relation as Eq.~\eqref{eq:intro-factorization}, now specialized to a general quasi-PDF scheme $S$, with the lightcone 
PDF fixed in the $\msbar$ scheme. The matching coefficient depends on the large momentum $|y|P^z$ carried by the parton entering 
the short-distance coefficient, rather than on the hadron momentum alone~\cite{Izubuchi:2018srq}.  In this paper the PDF on the 
right-hand side is kept in the $\msbar$ scheme, while $S$ specifies the renormalization prescription for the quasi-PDF.

To determine $C_S$ perturbatively, we replace the hadron by an external quark of momentum $p^z$.  The partonic factorization 
formula has the same convolution structure,
\begin{equation}
\tilde q_{q,{\text{ren}}}^{S}(x,p^z,\mu) = \int_{-1}^{1}\frac{dy}{|y|}\, C_S\left( \frac{x}{y}, \frac{\mu}{|y|p^z} \right)
q_{q,\msbar}(y,\mu),
\label{eq:matching-partonic}
\end{equation}
where the subscript $q$ denotes the external quark state.  At tree level,
\begin{equation}
q_q^{(0)}(x) = \tilde q_q^{(0)}(x) = C^{(0)}(x) = \delta(1-x).
\label{eq:tree-level}
\end{equation}
Expanding Eq.~\eqref{eq:matching-partonic} to one loop therefore gives~\cite{Xiong:2013bka,Izubuchi:2018srq}
\begin{equation}
C_S^{(1)}\left(x,\frac{\mu}{|p^z|}\right) = \tilde q_{q,{\text{ren}}}^{(1),S}(x,p^z,\mu) - q_{q,\msbar}^{(1)}(x,\mu).
\label{eq:matching-one-loop}
\end{equation}
The collinear IR contribution is common to the two terms on the right-hand side and cancels in the matching 
coefficient~\cite{Xiong:2013bka,Izubuchi:2018srq}.  We use dimensional regularization for this IR behavior on both sides of
Eq.~\eqref{eq:matching-one-loop}.

The scale dependence of the matching coefficient follows from
\begin{equation}
\frac{d q_{\msbar}}{d\ln\mu^2}=\Gamma_q\otimes q_{\msbar}, \qquad
\frac{d\tilde q_{\rm ren}^{S}}{d\ln\mu^2}=\Gamma_{\tilde q}^{S}\otimes\tilde q_{\rm ren}^{S}.
\label{eq:pdf-quasi-rg-general}
\end{equation}
The factorization formula then implies
\begin{equation}
\frac{d}{d\ln\mu^2}C_S = \Gamma_{\tilde q}^{S}\otimes C_S - C_S\otimes\Gamma_q,
\label{eq:matching-rg-general}
\end{equation}
and at one loop
\begin{equation}
\frac{d}{d\ln\mu^2}C_S^{(1)} = \Gamma_{\tilde q}^{(1),S} - \Gamma_q^{(1)}, \qquad \Gamma_q^{(1)}(x)=\asc P_{qq}(x),
\label{eq:matching-rg-one-loop-general}
\end{equation}
where $P_{qq}(x)$ is the one-loop quark-to-quark splitting function. Its explicit form is given in Eq.~\eqref{Pqq-def-tmc}.
These relations will be used below to identify the scale dependence associated with the TMC renormalization prescription.

\section{Ultraviolet structure in the transverse-momentum cutoff scheme} \label{sepdep}

In the TMC scheme, the transverse momentum is restricted by $k_\perp^2<\Lambda^2$, and the ultraviolet divergence is therefore 
expected to appear through the dependence on the cutoff $\Lambda$. There is, however, another region that has no counterpart for 
the ordinary quark PDF. At finite $p^z$, the momentum fraction $x$ of the quasi-PDF is not restricted to $0<x<1$. If the emitted 
gluon carries a large negative longitudinal momentum, the outgoing quark can carry longitudinal momentum larger than that of 
the parent quark, corresponding to $x>1$. Conversely, if the gluon carries a sufficiently large positive longitudinal momentum, 
the outgoing quark can have negative longitudinal momentum, corresponding to $x<0$. Hence $x$ can extend to $\pm\infty$. 
We must therefore examine whether the large-$|x|$ region gives rise to an ultraviolet divergence. If such a divergence exists and is 
intertwined with the ultraviolet contribution associated with the cutoff $\Lambda$, the two contributions must be disentangled in 
identifying the counterterm.

To see how these two ultraviolet contributions can be disentangled in general, it is useful to abstract the structure common to the 
one-loop diagrams computed in Sec.~\ref{matchone}. After the $k^0$ contour integration performed there, each diagram reduces 
to an integral over the transverse momentum of the generic form
\begin{equation} \label{sepi}
I(\Lambda,x)=\int_0^{\Lambda^2}d\kt\,f(\kt,x).
\end{equation}
A naive way to isolate the contribution associated with the transverse cutoff is
\begin{equation}
I(\Lambda,x) = \int_0^\infty d\kt\,f(\kt,x) - \int_{\Lambda^2}^\infty d\kt\,f(\kt,x).
\label{eq:naive-cutoff-decomposition}
\end{equation}
The second term contains the region above the transverse cutoff. Whether the divergence in Eq.~\eqref{sepi} is
logarithmic or a genuine power divergence, as in the tadpole diagram, its dependence on $\Lambda$ can only arise from
this term, since it alone carries $\Lambda$ as an explicit limit of integration. For a logarithmic divergence, both
terms separately carry the same $1/\euv$ pole from their shared large-$\kt$ tail. These poles cancel
between the two terms, and what survives is the $\Lambda$-dependent piece from the lower limit of the second term,
replacing the pole. For a genuine power divergence, dimensional regularization gives no pole in the first term at
all, and the $\Lambda$-dependence appears directly in the second. In either case, the second term correctly captures
the divergence associated with the cutoff. This decomposition, however, does not yet separate the two possible
sources of ultraviolet behavior identified above, since the first term still contains the complete large-$\kt$
region. What is needed instead is to separate the terms according to their large-$\kt$ behavior, retaining those terms whose t
ransverse-momentum integrals are not power suppressed in $\Lambda$. The remaining contribution can then be examined 
independently for ultraviolet behavior as $x\to\pm\infty$.

To disentangle the two regions, we define $f_{\text{asym}}(\kt,x)$ by retaining the terms in the large-$\kt$ expansion whose 
contributions after the transverse-momentum integration are not power suppressed in $\Lambda$. Adding and subtracting
$f_{\text{asym}}$ gives the exact identity
\begin{align} \label{decomp}
I(\Lambda,x) ={}& \int_0^{\Lambda^2}d\kt\,f_{\text{asym}}(\kt,x) + \int_0^\infty d\kt\, 
\bigl[f(\kt,x)-f_{\text{asym}}(\kt,x)\bigr] \nonumber\\
&- \int_{\Lambda^2}^\infty d\kt\, \bigl[f(\kt,x)-f_{\text{asym}}(\kt,x)\bigr].
\end{align}
The first term contains the ultraviolet contribution associated with the transverse cutoff. The second term is independent of 
$\Lambda$ and free of this transverse ultraviolet contribution, but its behavior as $x\to\pm\infty$ must still be examined. 
The last term is power suppressed in $\Lambda$. We therefore define
\begin{equation}
I_{\text{UV}}^{(\perp)}(\Lambda,x) \equiv \int_0^{\Lambda^2}d\kt\,f_{\text{asym}}(\kt,x), \qquad
I_{\text{rem}}(x) \equiv \int_0^\infty d\kt\, \bigl[f(\kt,x)-f_{\text{asym}}(\kt,x)\bigr],
\end{equation}
so that
\begin{equation} \label{schemesep}
I(\Lambda,x) = I_{\text{UV}}^{(\perp)}(\Lambda,x)+I_{\text{rem}}(x) +\mathcal O(1/\Lambda).
\end{equation}

Although $I_{\text{rem}}(x)$ is free of the ultraviolet contribution from large $k_\perp$, it may still be UV divergent as 
$x\to\pm\infty$. The transverse-momentum cutoff applies only to the transverse loop momentum and has no bearing on the 
momentum fraction $x$. We therefore continue to use dimensional regularization, already present for the collinear IR behavior, 
to examine this region. When extracting the $\Lambda$-dependent part, $\epsilon$ can be set to zero, since it does not enter 
the $x$-dependent terms there. It should not be set to zero from the outset, however, since it is exactly this regulator that exposes 
the ultraviolet behavior at $x=\pm\infty$. In the explicit calculation, the large-$|x|$ behavior of 
$I_{\text{rem}}(x)$ generates $1/\euv$ poles through the prescriptions on distributions at $x=\pm\infty$. Thus the two UV 
contributions appear respectively as $\Lambda$-dependent terms and $1/\euv$ boundary terms. The simultaneous appearance 
of $\Lambda$ and $\epsilon$ does not represent two regulators applied to the same ultraviolet region.

These two ultraviolet contributions determine the divergent part of the TMC counterterm but do not fix its finite part. 
The transverse-momentum cutoff does not preserve the Ward identity, so quark-number conservation is not automatic. For a 
non-anomalous symmetry, a Ward identity broken by the regularization prescription may be restored in the renormalized theory 
by an appropriate choice of counterterms, including their finite parts~\cite{Bonini:1993kt,Grassi:1999tp,Bochicchio:1985xa}. In the 
present calculation, we use this freedom to restore quark-number conservation at one loop and impose
\begin{equation}
\int_{-\infty}^{\infty}dx\,\tilde q_{\text{TMC,ren}}(x,\mu,p^z)=1.
\label{tmc-quark-number}
\end{equation}
The same condition was verified directly on the finite-cutoff result of Ref.~\cite{Xiong:2013bka}. That reference
similarly relies on the Ward identity to fix the one-loop wave-function renormalization constant, and reports that no
additional terms are needed to preserve the vector current at one loop. This presupposes that the finite freedom in
the counterterm has already been fixed consistently with the Ward identity, which is precisely the choice justified
explicitly in this paper through Eq.~\eqref{tmc-quark-number}.
 
\section{One-loop calculation in the TMC scheme} \label{matchone}

 \begin{figure}[b]
\centering
\includegraphics[width=\columnwidth]{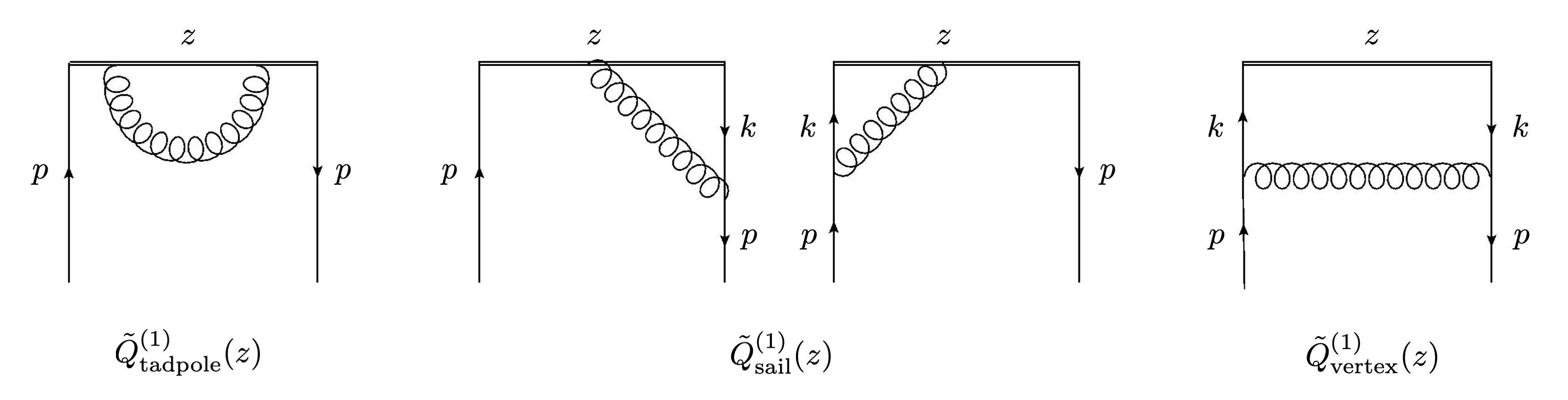}
\caption{ \label{fig1} 
The first tadpole diagram contains the linear divergence, the second and the third diagrams are called
the sail diagrams. The fourth diagram is the vertex correction. The Feynman diagram for the wave-function 
renormalization is not shown.}
\end{figure} 
 
We now apply the separation of Sec.~\ref{sepdep} to the explicit one-loop calculation. The one-loop calculation can be performed in 
any gauge, and we choose the Feynman gauge. The relevant Feynman diagrams are shown in Fig.~\ref{fig1} in coordinate space to 
display the nonlocal dependence of the quasi-PDF on the coordinate $z$. We perform the calculation in momentum space using the 
Fourier transform
\begin{equation}
\tilde{q} (x, \mu) = p_z \int \frac{dz}{2\pi} e^{i x z p_z} \tilde{Q} (z,\mu),
\end{equation}
because the transverse-momentum cutoff is implemented directly in momentum space. The Feynman rules for the quark
quasi-PDF operator with a straight spatial Wilson line are standard. We derived them independently and checked that
they agree with the results already available in the literature, and for brevity we do not reproduce the derivation
here. The one-loop contribution to the nonsinglet bare-quark quasi-PDF $\tilde Q(z,\mu)$ in the Feynman gauge can be
found in Ref.~\cite{Stewart:2017tvs} for $\gamma^z$ appearing in Eq.~\eqref{eq:quasi-bare-def} and in
Ref.~\cite{Izubuchi:2018srq} for $\gamma^t$. We use these coordinate-space expressions and take the Fourier transform to
obtain the quasi-PDF $\tilde q(x,\mu)$ in momentum-fraction space.

For each contribution we display the momentum-space integrand after the $k^0$ contour integration. At this stage the
transverse-momentum integration has not yet been performed, so this separation can be applied directly. We begin with the 
tadpole contribution, which gives the simplest illustration of the explicit $\Lambda$ dependence and also introduces the 
second-order prescription required by its $1/(1-x)^2$ behavior. We then turn to the sail and vertex diagrams. Their large-$k_\perp$ 
tails are power suppressed, but their large-$|x|$ behavior leads to the UV boundary terms at $x=\pm\infty$. The wave-function 
contribution is considered last before all one-loop terms are combined.

\subsection{Tadpole contribution}
 We begin with the tadpole contribution $\tilde{q}^{(1)}_{\text{tadpole}} (x)$, shown as the first diagram in
Fig.~\ref{fig1}, which arises from gluon exchange
between two points on the spatial Wilson line. The corresponding contribution vanishes for the lightcone PDF in the
Feynman gauge. The contraction of the two Wilson-line vertices is proportional to $z^2$ for the spatial Wilson line,
whereas it is proportional to $n^2=0$ for the lightlike Wilson line. The tadpole diagram provides the simplest example of the
power dependence on the transverse-momentum cutoff. When it is expressed as a distribution on the full line, its
$1/(1-x)^2$ behavior also requires a second-order prescription on distributions. We derive the required prescription
below when this behavior appears explicitly. In momentum space, the tadpole contribution is 
\begin{align} \label{tadpole}
\tilde{q}^{(1)}_{\text{tadpole}} (x)&= -ig^2 C_F  p_z\mud \int\ddk \frac{1}{k^2 +i0} \frac{1}{k_z^2} \Bigl[ \delta \bigl(
p_z (1-x)\bigr) - \delta \bigl( k_z - p_z (1-x) \bigr) \Bigr]  \\
&= -\frac{\alpha_s C_F}{4\pi} \dms \int_0^{\Lambda^2} d\kt \int_{-\infty}^{\infty} dk_z 
\frac{p_z(\kt)^{-\eps} }{k_z^2 \sqrt{k_z^2 + \kt}}
\Bigl[ \delta \bigl(p_z (1-x)\bigr) - \delta \bigl( k_z - p_z (1-x) \bigr) \Bigr], \nonumber
\end{align}
where $\tilde{\mu}^2$ denotes $\mu^2 e^{\gamma_E}/4\pi$. The second line is obtained by computing the contour integral in 
the complex $k_0$-plane, which is 
\begin{equation}
\int_{-\infty}^{\infty} \frac{dk_0}{k_0^2-k_z^2 -\kt +i0} = -\frac{i\pi}{\sqrt{k_z^2 +\kt}}. 
\end{equation}
In order to generate the appropriate distributions of $x$, let us consider the intermediate distribution
\begin{equation}
J_{\text{tadpole}} (x, \kt) = \int_{-\infty}^{\infty} dk_z  \frac{p_z}{k_z^2 \sqrt{k_z^2 + \kt}}
\Bigl[ \delta \bigl(p_z (1-x)\bigr) - \delta \bigl( k_z - p_z (1-x) \bigr) \Bigr], 
\end{equation}
which is the right-hand side of Eq.~\eqref{tadpole} without $\int d\kt (\kt)^{-\epsilon}$. It is regarded as a distribution in $x$ 
because it is singular at $x=1$. The action of the distribution $J_{\text{tadpole}} (x)$ on a smooth test function $\phi (x)$ is 
 \begin{align}
 \langle J_{\text{tadpole}} (x,\kt), \phi (x)\rangle &\equiv \int_{-\infty}^{\infty} dx J_{\text{tadpole}} (x,\kt) \phi (x)
 \nonumber \\
& = \int_{-\infty}^{\infty} dk_z \frac{1}{k_z^2 \sqrt{k_z^2 + \kt}} 
 \Bigl[ \phi (1) -\phi \Bigl( 1- \frac{k_z}{p_z}\Bigr) \Bigr]. 
 \end{align}
 By letting $x = 1-k_z/p_z$, it becomes
 \begin{align}
  &\langle J_{\text{tadpole}} (x,\kt), \phi (x)\rangle  = -\frac{1}{p_z} \int_{-\infty}^{\infty} dx 
   \frac{1}{(1-x)^2 \sqrt{(1-x)^2 p_z^2 +\kt}} \Bigl[ \phi (x) - \phi (1)\Bigr]   \\
  &=  -\frac{1}{p_z} \int_{-\infty}^{\infty} dx 
  \frac{1}{(1-x)^2 \sqrt{(1-x)^2 p_z^2 +\kt}} \Bigl[ \Bigl(\phi (x) - \phi (1) -(x-1) \frac{d\phi (1)}{dx}\Bigr) +
(x-1) \frac{d\phi (1)}{dx}\Bigr)   \Bigr].  \nonumber 
  \end{align}
In the second line we subtract and add $(x-1) d\phi (1)/dx$. The term proportional to $d\phi(1)/dx$ integrates to
zero because its integrand is odd under $(x-1) \rightarrow -(x-1)$. It can therefore be included without changing the
tadpole contribution, and the resulting expression supplies the second subtraction required to define the
$1/(1-x)^2$ singularity as a distribution. Including the integral $\int d\kt (\kt)^{-\epsilon}$, the kernel we consider
as a distribution is  
\begin{equation}
T (x,\eps,\Lambda) = \int_0^{\Lambda^2} d\kt (\kt)^{-\eps} J_{\text{tadpole}} (x,\kt)
=-\frac{1}{p_z}\frac{1}{(1-x)^2} \int_0^{\Lambda^2}\frac{d\kt (\kt)^{-\eps}}{\sqrt{(1-x)^2p_z^2+\kt}}.
\label{tadpole-kern}
\end{equation}
Now we extract the $\Lambda$-dependent part by taking the asymptotic form of the integrand. That is, we expand the integrand for 
large $\kt$ to the order it produces the power divergence in $\Lambda$, and it is given as
\begin{equation}
T_{\Lambda} (x,\eps, \Lambda) = -\frac{1}{p_z}\frac{1}{(1-x)^2} \int_0^{\Lambda^2}d\kt \,
\frac{(\kt)^{-\eps}}{\sqrt{\kt}} +\mathcal{O}\Bigl( \frac{1}{\Lambda} \Bigr) 
= -\frac{2\Lambda}{p_z} \frac{1}{(1-x)^2}+\mathcal{O}\Bigl( \frac{1}{\Lambda} \Bigr), 
\end{equation}
where the remaining terms are suppressed by $1/\Lambda$, and we put $\eps=0$ because $\eps$ does not
enter the $x$-dependent part. Therefore the action of $T_{\Lambda}$ on a test function $\phi (x)$ is
\begin{equation}
\left\langle T_{\Lambda}, \phi (x) \right\rangle = -\frac{2\Lambda}{p_z}  
\int_{-\infty}^{\infty} dx\,\frac{\phi(x)-\phi(1)-(x-1)\phi'(1)}{(1-x)^2}.
\end{equation}
The $\Lambda$-dependent part has a linear divergence in $\Lambda$ and is expressed as a double plus distribution of $x$,
which is
 \begin{equation}
T_{\Lambda} (x,\eps, \Lambda) = -\frac{2\Lambda}{p_z} \Bigl[\frac{1}{(1-x)^2}\Bigr]_{++(1)}^{[-\infty, \infty]},
\end{equation}
where we define the  double plus distribution centered at $x=1$ as
\begin{equation}
\left\langle \left[\frac{1}{(1-x)^2}\right]_{++(1)}^{[-\infty,\infty]}, \phi(x) \right\rangle \equiv
\int_{-\infty}^{\infty} dx\, \frac{\phi(x)-\phi(1)-(x-1)\phi'(1)}{(1-x)^2},
\label{eq:full-line-double-plus}
\end{equation}
for a test function  $\phi(x)$ sufficiently rapidly decreasing at infinity. This definition subtracts both the value and the 
first derivative of the test function at $x=1$.  It therefore fixes the endpoint-local part of the second-order plus distribution, 
so that no additional terms proportional to $\delta(1-x)$ or $d\delta(1-x)/dx$ are written separately. Other equivalent 
conventions differ only by such local endpoint terms.  Appendix~\ref{app:endpoint-distributions} explains this convention 
and its relation to those alternatives.
 
The remainder after subtracting the asymptotic $\Lambda$-dependent part is
\begin{align} \label{trem}
T_{\text{rem}} (x,\eps) &= -\frac{1}{p_z}\frac{1}{(1-x)^2} \int_0^{\infty} d\kt\,(\kt)^{-\eps} \left[
\frac{1}{\sqrt{(1-x)^2p_z^2+\kt}} -\frac{1}{\sqrt{\kt}} \right] \nonumber\\
&= -\frac{\Gamma(1-\epsilon)\Gamma (-\frac{1}{2}+\epsilon )}{\sqrt{\pi}}\, p_z^{-2\eps}\,
\frac{|1-x|^{1-2\eps}}{(1-x)^2}.
\end{align}
This expression is independent of the cutoff $\Lambda$, but we do not yet interpret it as a renormalized contribution 
to the quasi-PDF.  It is only the remnant left after the explicit large-$\kt$ term has been separated from the tadpole graph.  
The reason is that the quasi-PDF is a distribution on the whole real line. The endpoint singularity at $x=1$, and also the
boundary behavior at $x=\pm\infty$, should be treated only after all one-loop contributions are combined. We therefore
write the tadpole contribution as
\begin{equation} \label{tadres}
\tilde q_{\text{tadpole}}^{(1)}(x,\Lambda,\mu,p_z) = \tilde q_{\text{tadpole}}^{(1),\Lambda}(x,\Lambda,\mu,p_z)
+ \tilde q_{\text{tadpole}}^{(1),\text{rem}}(x,\mu,p_z),
\end{equation}
where
\begin{equation} \label{tadpole-Lambda}
\tilde q_{\text{tadpole}}^{(1),\Lambda}(x,\Lambda,\mu,p_z) = \asc \frac{\Lambda}{p_z} \left[ \frac{1}{(1-x)^2}
\right]_{++(1)}^{[-\infty,\infty]} ,
\end{equation}
Multiplying by the overall prefactor $-\frac{\alpha_s C_F}{4\pi}\dms$, giving
\begin{equation} \label{tadpole-rem}
\tilde q_{\text{tadpole}}^{(1),\text{rem}}(x,\mu,p_z) = \frac{1}{2}\asc
\Bigl(\frac{\mu^2 e^{\gamma_E}}{p_z^2}\Bigr)^{\eps}
\frac{\Gamma (-\frac{1}{2}+\epsilon )}{\sqrt{\pi}}\,
\frac{|1-x|^{1-2\eps}}{(1-x)^2}.
\end{equation}
 
The treatment of the linear divergence depends on which matching problem is being considered. In the matching
between the renormalized quasi-PDF and the lightcone PDF, the situation is asymmetric. The lightcone PDF has no
spatial Wilson line and hence no linear divergence, while the TMC quasi-PDF does. Since this contribution has no
counterpart to match against, it must first be assigned to the quasi-PDF counterterm.  After this
power subtraction is made, the remaining quasi-PDF can be matched to the lightcone PDF through the usual logarithmic 
and finite terms. There is a second matching problem involving the quasi-PDF and the spatial correlation function  in 
lattice QCD.  In that case both quantities contain a spatial Wilson line, and both can contain a linear divergence. The 
matching between them should be formulated after choosing compatible prescriptions for the power divergence in the 
two descriptions. Once the linear divergence is treated consistently, the remaining difference is governed by logarithmic 
cutoff dependence and finite terms.  

In the present paper we focus on the first matching problem.  We therefore separate the linearly divergent 
contribution as part of the TMC counterterm before extracting the logarithmic structure relevant to the matching between
the quasi-PDF and the PDF. To be more specific on the treatment of the linear divergence, we can employ the idea 
in lattice QCD.  In lattice calculations, the corresponding Wilson-line power divergence is removed by mass
renormalization~\cite{Chen:2016fxx,Green:2017xeu}.  This treatment is consistent with the multiplicative renormalizability 
of the underlying nonlocal quasi-parton operators~\cite{Ishikawa:2017faj,Li:2018tpe}.  The same issue appears in the TMC 
scheme because the quasi-PDF contains a linearly divergent contribution that has no counterpart in the lightcone PDF.
Therefore a corresponding scheme for the linear divergence must be included in the definition of the renormalized quasi-PDF.
In coordinate space this subtraction may be written schematically as
\begin{equation}
  O_R(z,\mu)   =   e^{+\delta m |z|}   Z_{\text{log}}^{-1}(\Lambda,\mu)   O_{\text{bare}}(z).
  \label{cormass}
\end{equation}
The factor $e^{+\delta m |z|}$ represents the mass renormalization that removes the linear divergence, taking this
separate form because the divergence is assigned to the TMC counterterm on its own. The remaining renormalization,
contained in $Z_{\text{log}}$, removes the logarithmic cutoff dependence and will be developed as the calculation
proceeds. In the present TMC calculation, Eq.~\eqref{cormass} should be understood as the coordinate-space
counterpart of this counterterm assignment, in analogy with the mass renormalization used in lattice calculations.

\subsection{Sail and vertex diagrams}
The tadpole diagram illustrates how the explicit $\Lambda$ dependence is isolated from the large-$k_\perp$ behavior. We now 
turn to the sail and vertex diagrams, for which the large-$k_\perp$ asymptotic terms are power suppressed. These diagrams do
not generate nonvanishing dependence on $\Lambda$ as $\Lambda\to\infty$, but their large-$|x|$ behavior becomes
important when the result is defined as a distribution on the full line.

The sail diagrams correspond to the two diagrams in the middle of Fig.~\ref{fig1}. The sail diagram in momentum space 
is given as
\begin{align} \label{sail}
\tilde{q}_{\text{sail}}^{(1)} (x,\mu) &= -\frac{\as C_F}{4\pi} \dms\int_0^1 du \int_0^{\Lambda^2} d\kt (\kt)^{-\eps} 
\int_{-\infty}^{\infty} dk_z \frac{2(1-u) p_z -k_z}{(k_z^2 + \kt)^{3/2} (k_z +up_z)}  \nonumber \\
&\times p_z \Bigl[ \delta \bigl(p_z (1-x)\bigr) - \delta \bigl( k_z - p_z (1-u-x)\bigr) \Bigr],
\end{align}
where $u$ is the Feynman parameter to combine the denominator in the original integrand, and the contour integral on 
the complex $k_0$-plane is performed.

Consider the following intermediate distribution without $\int d\kt (\kt)^{-\eps}$
\begin{equation}
J_{\text{sail}} (x, \kt) = \int_{-\infty}^{\infty}  dk_z \frac{2(1-u) p_z -k_z}{(k_z^2 + \kt)^{3/2} (k_z +up_z)}   
p_z \Bigl[ \delta \bigl(p_z (1-x)\bigr) - \delta \bigl( k_z - p_z (1-u-x)\bigr) \Bigr].
\end{equation}
The action of the kernel to  a smooth test function $\phi (x)$ is
\begin{align}
&\langle J_{\text{sail}} (x, \kt), \phi (x) \rangle = \int_{-\infty}^{\infty}  dk_z 
\frac{2(1-u) p_z -k_z}{(k_z^2 + \kt)^{3/2} (k_z +up_z)} 
\Bigl[ \phi (1) - \phi \Bigl( 1- u-\frac{k_z}{p_z} \Bigr) \Bigr]  \nonumber \\
& =- \int_{-\infty}^{\infty} dx \frac{1-u+x}{(1-x) \bigl[ (1-u-x)^2 p_z^2 + \kt\bigr]^{3/2}} \Bigl[ \phi (x) -\phi (1)\Bigr],
\end{align}
where we change the variable $ x = 1-u-k_z/p_z$ to obtain the last expression. Here again we can extract the asymptotic 
behavior for large $\kt$ to find a divergent term. But the leading asymptotic term scales as $\sim (\kt)^{-3/2-\eps}$, and 
it is suppressed by powers of $1/\Lambda$. Therefore the expression for the sail diagram in Eq.~\eqref{sail} itself 
corresponds to the remainder, independent of the transverse-momentum cutoff $\Lambda$. The whole kernel for the
sail diagram by including the integral $\int d\kt (\kt)^{-\eps}$ is given as
\begin{equation}
S_{\text{rem}} (x,\eps) = -2 \Gamma (1-\eps) \frac{\Gamma (1/2+\eps)}{\sqrt{\pi}} p_z^{-2\eps} \int_0^1 du
\frac{1-u+x}{1-x} |1-u-x|^{-1-2\eps}.
\end{equation}
Therefore the contribution of the sail diagrams to the remainder is given as
\begin{align}
\tilde{q}_{\text{sail}}^{(1),\text{rem}} (x,\mu)&= \frac{\alpha_s C_F}{2\pi}  
\Bigl( \frac{\mu^2 e^{\gamma_E}}{p_z^2}\Bigr)^{\eps}
\frac{\Gamma (1/2+\eps)}{\sqrt{\pi}}  \int_0^1 du \frac{1-u+x}{1-x} |1-u-x|^{-1-2\eps}.
\end{align}
 
The vertex diagram in coordinate space is the last diagram in Fig.~\ref{fig1}. 
After the $k_0$ contour integral, the vertex diagram in momentum space is given as
\begin{equation}
\tilde{q}_{\text{vertex}}^{(1)} = \frac{\as C_F}{4\pi}  \dms  (1-\eps)  \int_0^1 du (1-u) \int dk_z  \int d\kt 
\frac{(\kt)^{-\eps}p_z \delta (k_z - (x-u) p_z) }{\bigl[ (x-u)^2 p_z^2 + \kt\bigr]^{3/2}},
\end{equation}
where $u$ is also the Feynman parameter to combine the denominators in the original integrand. The asymptotic 
part is proportional to $(\kt)^{-3/2-\eps}$, so $\tilde{q}_{\text{vertex}}^{(1)}$ is
itself the remainder, with no explicit $\Lambda$ dependence. After performing the momentum integration,
the contribution of the vertex diagram to the remainder is given as
\begin{align}
\tilde{q}_{\text{vertex}}^{(1),\text{rem}} (x,\mu)&= \frac{\as C_F}{2\pi}  
\Bigl( \frac{\mu^2 e^{\gamma_E}}{p_z^2}\Bigr)^{\eps}
\frac{\Gamma (1/2+\eps)}{\sqrt{\pi}} (1-\eps) \int_0^1 du (1-u) |x-u|^{-1-2\eps}.
\end{align}

It is useful to contrast this result with the calculation in dimensional regularization by Izubuchi \textit{et al.}~\cite{Izubuchi:2018srq}. 
Their coordinate-space sail contribution contains an explicit combination $1/\euv-1/\eir$. In the present calculation, however, the 
large-$k_\perp$ behavior of the sail contribution after the $k^0$ contour integration is proportional to $(\kt)^{-3/2-\epsilon}$. 
The region above the transverse-momentum cutoff is therefore power suppressed, and no $\ln(\Lambda^2/\mu^2)$ term survives 
as $\Lambda\to\infty$.

This provides a direct example of the separation discussed in Sec.~\ref{sepdep}. A UV pole appearing in dimensional
regularization cannot in general be identified diagram by diagram with logarithmic dependence on the transverse-momentum
cutoff. The latter is determined by the large-$k_\perp$ asymptotic behavior. The sail and vertex diagrams have no 
$\Lambda$-dependent contribution, while their large-$|x|$ behavior remains in the remainder. As discussed
by Izubuchi \textit{et al.} in their momentum-fraction representation and demonstrated explicitly in their treatment of
the distributions at infinity, the divergence at $x=\pm\infty$ can generate UV poles. We will see this boundary UV
contribution explicitly below when the remainder is defined as a distribution on the full line.

\subsection{Wave-function renormalization}\label{sec:wf-renormalization-cutoff-split}

Before combining all one-loop contributions into a distribution on the full line, we need the
wave-function contribution. In the TMC scheme it is
\begin{equation} \label{wfren}
\tilde{q}_{\text{wf}}^{(1)} (x,\Lambda,\mu)  = -\frac{\as C_F}{4\pi}  \left(  -\frac{1}{\epsilon_{\text{IR}}}  +1
+\ln\frac{\Lambda^2}{\mu^2}  \right)\delta(1-x).
\end{equation}
The logarithmic term $\ln(\Lambda^2/\mu^2)$ is the remaining explicit cutoff dependence at one loop and contributes to
the anomalous dimension. The $1/\eir$ pole belongs to the remainder and is needed for the cancellation of the collinear
singularity in matching with the PDF. The finite constant is not fixed by removing the UV divergence, and we leave its
assignment open until all one-loop contributions have been combined. We therefore separate only the explicit
$\Lambda$ dependence at this stage. The superscript ``0'' on the remainder indicates that its finite constant is not
yet fixed, unlike the sail and vertex remainders, which do not need that distinction.
\begin{equation}
\tilde{q}_{\text{wf}}^{(1)}(x,\Lambda,\mu) = \tilde{q}_{\text{wf}}^{(1),\Lambda}(x,\Lambda,\mu)
+ \tilde{q}_{\text{wf}}^{(1),\text{rem},0}(x,\mu),
\end{equation}
where
\begin{align}\label{wfrenpart}
\tilde{q}_{\text{wf}}^{(1),\Lambda}(x,\Lambda,\mu)
&= -\frac{\as C_F}{4\pi}\ln\frac{\Lambda^2}{\mu^2}\,\delta(1-x), \nonumber\\
\tilde{q}_{\text{wf}}^{(1),\text{rem},0}(x,\mu)
&= \frac{\as C_F}{4\pi}\left(\frac{1}{\eps_{\text{IR}}}-1\right)\delta(1-x).
\end{align}
The finite term will be assigned only after the boundary UV contribution has also been identified.
Having evaluated all one-loop diagrams, we now combine their remainders and complete the definition of the quasi-PDF as
a distribution on the full line.

\subsection{Distributions on the full line from the quasi-PDF at one loop}

We now collect the one-loop contributions after the explicit $\Lambda$-dependent terms have been separated.  We first
display the result at finite $x$. This intermediate remnant should not be identified with the fully renormalized 
quasi-PDF yet because finally we need a distribution on the whole real line.  The reason is that the outer regions have 
large-$|x|$ tails whose distribution generates UV boundary poles at $x=\pm\infty$.  We therefore first write the remnant 
at finite $x$, and then include the boundary contribution in the TMC counterterm before giving the
renormalized quasi-PDF on the full line.

At this stage we write
\begin{equation}
\tilde{q}_{\text{bare}}^{(1)}(x,\Lambda,\mu,p_z) = \tilde{q}_{\text{TMC,ct}}^{(1),\Lambda}
(x,\Lambda,\mu,p_z) + \tilde{q}_{\text{rem}}^{(1),0}(x,\mu,p_z),
\label{tmc-first-decomp-explicit}
\end{equation}
where $\tilde q_{\text{TMC,ct}}^{(1),\Lambda}$ contains only the terms with explicit dependence on the
transverse-momentum cutoff $\Lambda$. At one loop these are the linearly divergent tadpole contribution and the logarithmic
wave-function contribution,
\begin{align}
\tilde{q}^{(1),\Lambda}_{\text{TMC,ct}}(x,\Lambda,\mu,p_z) &= 
\tilde{q}_{\text{tadpole}}^{(1),\Lambda}(x,\Lambda,\mu,p_z) +
\tilde{q}_{\text{wf}}^{(1),\Lambda}(x,\Lambda,\mu) \nonumber\\
&= \asc \left[ \frac{\Lambda}{p_z} \left[\frac{1}{(1-x)^2}\right]_{++(1)}^{[-\infty,\infty]} -
\frac{1}{2}\delta(1-x)\ln\frac{\Lambda^2}{\mu^2} \right].
\label{tmc-counterterm-Lambda}
\end{align}
No finite counterterm has been chosen at this stage. The remaining part is
\begin{align}
\tilde{q}^{(1),0}_{\text{rem}}(x,\mu,p_z) &= \tilde{q}^{(1),\text{rem}}_{\text{tadpole}}(x,\mu,p_z)
+ \tilde{q}_{\text{vertex}}^{(1),\text{rem}}(x,\mu,p_z) 
+ \tilde{q}_{\text{sail}}^{(1),\text{rem}}(x,\mu,p_z) + \tilde{q}_{\text{wf}}^{(1),\text{rem},0}(x,\mu)
\nonumber \\
&= \asc \Bigl(\frac{\mu^2 e^{\gamma_{\text{E}}}}{p_z^2}\Bigr)^\eps \frac{\Gamma(1/2+\eps)}{\sqrt{\pi}}
\Biggl[ \frac{|1-x|^{1-2\eps}}{2(-1/2+\eps)(1-x)^2} \nonumber\\
& +(1-\eps)\int_0^1 du\,(1-u)|x-u|^{-1-2\eps} +\int_0^1 du\,\frac{1-u+x}{1-x}|1-u-x|^{-1-2\eps}
\Biggr] \nonumber\\
&\quad + \asc \frac{1}{2}\left(\frac{1}{\eps_{\text{IR}}}-1\right)\delta(1-x).
\label{qrem-unexpanded}
\end{align}
Expanding this expression at finite $x$, one obtains
\begin{align} \label{qrem-finite-x}
\tilde{q}^{(1),0}_{\text{rem}}(x,\mu,p_z)  &= \asc \left[ -\frac{1}{\eps_{\text{IR}}} P_{qq}(x)\theta(x)\theta(1-x)
+\left( \frac{3}{2}\ln\frac{\mu^2}{4p_z^2} +2 \right)\delta(1-x) \right] \\
&\quad + \asc \left\{
\begin{array}{ll}
\displaystyle \left[ \frac{1+x^2}{1-x}\ln\frac{x}{x-1} +1+\frac{3}{2x} \right]_{+(1)}^{[1,\infty]} -\frac{3}{2x}, & x>1,
\\[2.0ex]
\displaystyle \left[ \frac{1+x^2}{1-x} \ln\frac{4x(1-x)p_z^2}{\mu^2} -\frac{x(1+x)}{1-x} \right]_{+(1)}^{[0,1]}, & 0<x<1,
\\[2.0ex]
\displaystyle \left[ \frac{1+x^2}{1-x}\ln\frac{1-x}{-x} -1+\frac{3}{2(1-x)} \right]_{+(1)}^{[-\infty,0]} -\frac{3}{2(1-x)}, & x<0 .
\end{array}
\right. \nonumber
\end{align}
Here the quark splitting function $P_{qq} (x)$ is 
\begin{equation}
P_{qq}(x) = \left[\frac{1+x^2}{1-x}\right]_{+(1)}^{[0,1]} = 2\left[\frac{1}{1-x}\right]_{+(1)}^{[0,1]} -(1+x) 
+\frac{3}{2}\delta(1-x).
\label{Pqq-def-tmc}
\end{equation}
The pole term in Eq.~\eqref{qrem-finite-x} is the collinear infrared singularity common to the quasi-PDF and the 
lightcone PDF.  It is left in the remnant and cancels in the matching coefficient. 

Note that we have rearranged some functions of $x$ for $x>1$ and $x<0$. Those combinations in the brackets fall as $1/x^2$ 
at infinity. By this separation, the singularity at infinity no longer needs to be considered for these combinations, and only
the remaining terms need to be examined at large $|x|$. As $x \to  \pm\infty$, the following 
functions  behave as
\begin{align}
\frac{1+x^2}{1-x}\ln\frac{x}{x-1}+1 &= -\frac{3}{2x} +\mathcal{O}(x^{-2}), \qquad x\to+\infty, \nonumber\\
\frac{1+x^2}{1-x}\ln\frac{1-x}{-x}-1 &= -\frac{3}{2(1-x)} +\mathcal{O}(x^{-2}), \qquad x\to-\infty .
\label{outer-tail-asymptotic}
\end{align}
We therefore add and subtract $3/(2x)$ for $x>1$ and $3/[2(1-x)]$ for $x<0$ in Eq.~\eqref{qrem-finite-x}. 
The terms inside the square brackets then fall as $1/x^2$ at infinity and define ordinary plus distributions
with respect to the endpoint $x=1$.  The subtracted tails are kept separately, because their expression of the distributions 
at infinity produces the boundary UV poles, which are absorbed into the TMC counterterm.

The remaining tail terms in Eq.~\eqref{qrem-finite-x} must now be expanded as distributions at $x=\pm\infty$.  The 
corresponding boundary contribution to the TMC counterterm is
\begin{align}
\tilde q_{\text{TMC,ct}}^{(1),\infty}(x) &= -\asc \,\frac{3}{2}\cdot\frac{1}{2\euv}
\left[ \frac{1}{x^2}\delta^+\!\left(\frac{1}{x}\right) + \frac{1}{(1-x)^2} \delta^+\!\left(\frac{1}{1-x}\right) \right].
\label{tmc-counterterm-infinity}
\end{align}
The first term is supported at $x=+\infty$, while the second is supported at $x=-\infty$.  The superscript $+$ indicates 
that the argument of the delta function approaches zero from the positive side.  The factor $1/2$ in the pole coefficient 
comes from the expansion of the regulated tails $x^{-1-2\epsilon}$ and $(1-x)^{-1-2\epsilon}$.  The same expansion also
leaves the finite boundary contribution
\begin{align}
\tilde q_{\text{rem}}^{(1),\infty,\text{fin}}(x) ={}& -\asc \frac{1}{2} \left( \frac{3}{2}\ln\frac{\mu^2}{4p_z^2}+\frac{5}{2} \right)
\nonumber\\
&\quad \times \left[ \frac{1}{x^2}\delta^+\!\left(\frac{1}{x}\right) + \frac{1}{(1-x)^2}\delta^+\!\left(\frac{1}{1-x}\right) \right].
\label{qrem-finite-infinity}
\end{align}
Both the UV pole and the finite boundary term are derived in
Appendix~\ref{app:infinity-distribution-form}.

The $1/\euv$ poles in Eq.~\eqref{tmc-counterterm-infinity} realize the boundary UV contribution anticipated in
Sec.~\ref{sepdep}. Although the transverse-momentum UV region has been regulated by $\Lambda$, the dimensional regulator
has been retained in Eq.~\eqref{qrem-unexpanded} in order to define the resulting quasi-PDF as a distribution on the full
line. The $\epsilon$-dependent large-$|x|$ tails then generate the $1/\euv$ terms when expanded as distributions at
$x=\pm\infty$. These poles therefore originate from the boundaries of the momentum-fraction distribution rather than from
the transverse-momentum region regulated by $\Lambda$.

At this point the UV contributions have been identified. Their removal does not determine the remaining finite part of
the counterterm. As explained in Sec.~\ref{sepdep}, we use this remaining freedom to restore quark-number conservation,
broken by the transverse-momentum cutoff~\cite{Bonini:1993kt,Grassi:1999tp,Bochicchio:1985xa}. Here we show explicitly that this 
can be done at one loop. We choose
\begin{equation}
\tilde q_{\text{TMC,ct}}^{(1),\text{fin}}(x)=-\asc\frac{1}{2}\delta(1-x),
\label{tmc-counterterm-finite}
\end{equation}
so that the renormalized nonsinglet quasi-PDF preserves the quark number. This is a choice of the finite TMC prescription,
not a condition required for removing the UV divergences. Our explicit construction is restricted to one loop, and we do
not assume that the same prescription extends to all orders.

The full TMC counterterm is therefore
\begin{equation}
\tilde q_{\text{TMC,ct}}^{(1)} = \tilde q_{\text{TMC,ct}}^{(1),\Lambda} + \tilde q_{\text{TMC,ct}}^{(1),\infty}
+ \tilde q_{\text{TMC,ct}}^{(1),\text{fin}},
\label{tmc-counterterm-split}
\end{equation}
or explicitly
\begin{align}
\tilde{q}^{(1)}_{\text{TMC,ct}}(x,\Lambda,\mu,p_z) &= \asc \Biggl[ \frac{\Lambda}{p_z}
\left[\frac{1}{(1-x)^2}\right]_{++(1)}^{[-\infty,\infty]} - \frac{1}{2}\delta(1-x)\ln\frac{\Lambda^2}{\mu^2} \nonumber\\
&\hspace{2.0cm}
-\frac{1}{2}\delta(1-x)
-\frac{3}{4\euv} \frac{1}{x^2}\delta^+\!\left(\frac{1}{x}\right) - \frac{3}{4\euv} \frac{1}{(1-x)^2}
\delta^+\!\left(\frac{1}{1-x}\right) \Biggr].
\label{tmc-counterterm}
\end{align}
The linearly divergent term and the logarithmic wave-function term in Eq.~\eqref{tmc-counterterm} come from the
explicit cutoff dependence. The finite $-\delta(1-x)/2$ term specifies the finite TMC scheme chosen here and preserves
quark-number conservation. The last two terms have a different origin. They are UV boundary contributions of the
momentum-fraction distribution on the full line and are exposed by dimensional regularization. The term with
$\delta^+(1/x)$ comes from the $x\to+\infty$ tail, while the term with $\delta^+(1/(1-x))$ comes from the
$x\to-\infty$ tail. Thus the complete counterterm displays separately the transverse UV contribution, the boundary UV
contribution, and the finite scheme choice.

After the boundary UV contribution and the finite prescription have been specified, the renormalized quasi-PDF is
obtained as a distribution on the full line,
\begin{equation}
\tilde q_{\text{rem}}^{(1),0} = \tilde q_{\text{TMC,ct}}^{(1),\infty}
+ \tilde q_{\text{TMC,ct}}^{(1),\text{fin}} + \tilde q_{\text{TMC,ren}}^{(1)},
\label{qrem-second-decomp}
\end{equation}
Since $\tilde q_{\text{bare}}^{(1)} = \tilde q_{\text{TMC,ct}}^{(1),\Lambda} + \tilde q_{\text{rem}}^{(1),0}$
from Eq.~\eqref{tmc-first-decomp-explicit}, and the full counterterm is $\tilde q_{\text{TMC,ct}}^{(1)} =
\tilde q_{\text{TMC,ct}}^{(1),\Lambda} + \tilde q_{\text{TMC,ct}}^{(1),\infty} + \tilde q_{\text{TMC,ct}}^{(1),\text{fin}}$
from Eq.~\eqref{tmc-counterterm-split}, Eq.~\eqref{qrem-second-decomp} is equivalent to
\begin{equation}
\tilde q_{\text{bare}}^{(1)} = \tilde q_{\text{TMC,ct}}^{(1)} + \tilde q_{\text{TMC,ren}}^{(1)}.
\label{tmc-decomp-explicit}
\end{equation}
The final result is
\begin{align}
\tilde q_{\text{TMC,ren}}^{(1)}(x,\mu,p_z) &= \asc \left[ -\frac{1}{\eps_{\text{IR}}} P_{qq}(x)\theta(x)\theta(1-x)
+ \left( \frac{3}{2}\ln\frac{\mu^2}{4p_z^2} +\frac{5}{2} \right)\delta(1-x) \right] \nonumber\\
&\quad + \asc \Biggl\{ \left[ \frac{1+x^2}{1-x}\ln\frac{x}{x-1} +1+\frac{3}{2x} \right]_{+(1)}^{[1,\infty]} -\frac{3}{2}
\left(\frac{1}{x}\right)_{+(\infty)}^{[1,\infty]} \nonumber\\
&\qquad + \left[ \frac{1+x^2}{1-x} \ln\frac{4x(1-x)p_z^2}{\mu^2} -\frac{x(1+x)}{1-x} \right]_{+(1)}^{[0,1]} \nonumber\\
&\qquad + \left[ \frac{1+x^2}{1-x}\ln\frac{1-x}{-x} -1+\frac{3}{2(1-x)} \right]_{+(1)}^{[-\infty,0]} -\frac{3}{2}
\left(\frac{1}{1-x}\right)_{+(-\infty)}^{[-\infty,0]} \Biggr\} \nonumber\\
&\quad -\asc \frac{1}{2} \left( \frac{3}{2}\ln\frac{\mu^2}{4p_z^2}+\frac{5}{2} \right) \left[
\frac{1}{x^2}\delta^+\!\left(\frac{1}{x}\right) + \frac{1}{(1-x)^2}\delta^+\!\left(\frac{1}{1-x}\right) \right].
\label{finalqrem}
\end{align}
Here and below, the renormalized quasi-PDF refers to the quasi-PDF after UV renormalization.  The collinear IR pole remains in the 
partonic matrix element and cancels against the corresponding pole in the PDF when the matching coefficient is constructed.
The finite boundary term in Eq.~\eqref{finalqrem} is required by the same distributional expansion that produces the boundary 
UV pole in Eq.~\eqref{tmc-counterterm-infinity}.  The explicit delta functions in the finite part occur in the combination
\begin{equation}
\left(\frac{3}{2}\ln\frac{\mu^2}{4p_z^2}+\frac{5}{2}\right) \left[
\delta(1-x)-\frac{1}{2}\frac{1}{x^2}\delta^+\!\left(\frac{1}{x}\right)
-\frac{1}{2}\frac{1}{(1-x)^2}\delta^+\!\left(\frac{1}{1-x}\right) \right].
\label{tmc-delta-number-combination}
\end{equation}
The three distributions in brackets have zeroth moments $1$, $-1/2$, and $-1/2$, respectively, so their sum vanishes.
All the remaining finite terms are written with plus prescriptions and also have vanishing zeroth moment.  The
renormalized quasi-PDF therefore satisfies the quark-number condition in Eq.~\eqref{tmc-quark-number}.

The boundary delta functions in the counterterm do not contribute to the matching coefficient at finite $x$, but they are 
part of the UV renormalization of the quasi-PDF as a distribution on the whole real line.  Consequently they also enter 
the anomalous dimension, to be discussed in Section~\ref{rg-tmc}.

To complete matching, we need the lightcone PDF.  The one-loop bare PDF can
be separated into the UV counterterm and the IR-divergent part as
\begin{align}
q^{(1)}(x,\euv) &= \asc \frac{1}{\euv} P_{qq}(x)\theta(x)\theta(1-x), \nonumber\\
q^{(1)}(x,\eir) &= -\asc \frac{1}{\eir} P_{qq}(x)\theta(x)\theta(1-x).
\label{pdf-uv-ir}
\end{align}
The UV pole in $q^{(1)}(x,\euv)$ is removed by the $\msbar$ renormalization of the lightcone PDF, while the IR pole in
$q^{(1)}(x,\eir)$ is the collinear singularity which enters matching. This IR pole cancels the corresponding pole in the 
renormalized quasi-PDF.

Using the renormalized quasi-PDF on the full line in Eq.~\eqref{finalqrem}, the one-loop matching coefficient is obtained from
\begin{equation}
C_{\text{TMC}}^{(1)}(x,\mu,p_z) = \tilde q_{\text{TMC,ren}}^{(1)}(x,\mu,p_z) - q^{(1)}(x,\eir).
\label{matching-coeff-def}
\end{equation}
The boundary UV poles at $x=\pm\infty$ have already been absorbed into the TMC counterterm in
Eq.~\eqref{tmc-counterterm}.  They therefore do not appear in the finite matching coefficient.  The finite boundary
delta functions, however, remain part of the matching coefficient when it is written as a distribution on the full line.
The matching coefficient is
\begin{align} \label{coefftmc}
C_{\text{TMC}}^{(1)}(x,\mu,p_z) &= \asc \left[ \left( \frac{3}{2}\ln\frac{\mu^2}{4p_z^2} +\frac{5}{2} \right)\delta(1-x)
\right] \nonumber\\
&\quad + \asc \Biggl\{ \left[ \frac{1+x^2}{1-x}\ln\frac{x}{x-1} +1+\frac{3}{2x} \right]_{+(1)}^{[1,\infty]} -\frac{3}{2}
\left(\frac{1}{x}\right)_{+(\infty)}^{[1,\infty]} \nonumber\\
&\qquad + \left[ \frac{1+x^2}{1-x} \ln\frac{4x(1-x)p_z^2}{\mu^2} -\frac{x(1+x)}{1-x} \right]_{+(1)}^{[0,1]} \nonumber\\
&\qquad + \left[ \frac{1+x^2}{1-x}\ln\frac{1-x}{-x} -1+\frac{3}{2(1-x)} \right]_{+(1)}^{[-\infty,0]} -\frac{3}{2}
\left(\frac{1}{1-x}\right)_{+(-\infty)}^{[-\infty,0]} \Biggr\} \nonumber\\
&\quad -\asc \frac{1}{2} \left( \frac{3}{2}\ln\frac{\mu^2}{4p_z^2}+\frac{5}{2} \right)
\left[ \frac{1}{x^2}\delta^+\!\left(\frac{1}{x}\right) + \frac{1}{(1-x)^2}\delta^+\!\left(\frac{1}{1-x}\right) \right].
\end{align}
 
It is instructive to compare Eq.~\eqref{coefftmc} with the one-loop $\msbar$ quasi-PDF matching coefficient on the full
line in Ref.~\cite{Izubuchi:2018srq}.  After the TMC counterterm is defined by the prescription above, the result agrees
with the dimensional-regularization result, including the finite coefficient of $\delta(1-x)$ and the finite boundary
delta functions at $x=\pm\infty$.  This agreement was not imposed in determining the UV counterterm.  The finite
wave-function term is fixed by the quark-number condition, while the boundary terms follow independently from the
regulated large-$|x|$ expansion.  The agreement with Ref.~\cite{Izubuchi:2018srq} therefore provides a consistency check
of the complete TMC prescription.

The matching coefficient in Eq.~\eqref{coefftmc} is a distribution on the full line and therefore includes the finite
boundary terms at $x=\pm\infty$.  These terms do not contribute, however, to the factorization formula relating the
hadronic quasi-PDF to the lightcone PDF, since the convolution probes finite momentum fractions.  They must nevertheless
be retained in the complete matching coefficient.

 \section{Renormalization-group behavior in the TMC scheme}
\label{rg-tmc}

We now discuss the RG evolution of the renormalized quasi-PDF in the TMC scheme and the matching
coefficient.  The Wilson-line power divergence has already been removed separately and does not enter the RG evolution
discussed here.  The remaining UV renormalization contains the logarithmic contribution at finite $x$ together with the
boundary contributions at $x=\pm\infty$.  Consequently, the anomalous dimension of the quasi-PDF is itself a distribution
on the full line.  We do not attempt to solve the evolution equation explicitly.

The finite term added to the wave-function counterterm by the quark-number condition in
Eq.~\eqref{tmc-quark-number} is independent of $\mu$ and does not contribute to the anomalous dimension.  The
scale-dependent delta-function terms in the renormalized quasi-PDF occur in the combination shown in
Eq.~\eqref{tmc-delta-number-combination}.  These delta-function terms arise entirely from the ultraviolet counterterm,
in contrast to the collinear structure of the quasi-PDF at finite $x$ away from $x=1$, which is common to any
renormalization scheme and is carried separately by $\Gamma_q$ in the matching relation below.  The one-loop
ultraviolet anomalous dimension associated with the counterterm can therefore be written as
\begin{equation}
\Gamma_{\tilde q}^{\text{TMC}}(\as,x) = \Gamma_{\tilde q}^{\text{TMC,fin}}(\as,x)
+ \Gamma_{\tilde q}^{\mathrm{TMC,\infty}}(\as,x),
\label{gamma-tmc-split}
\end{equation}
with
\begin{equation}
\Gamma_{\tilde q}^{\text{TMC,fin}}(\as,x) = \asc\,\frac{3}{2}\delta(1-x) +\mathcal{O}(\as^2),
\label{gamma-tmc-finite}
\end{equation}
and
\begin{align}
\Gamma_{\tilde q}^{\mathrm{TMC,\infty}}(\as,x) = -\asc\,\frac{3}{4} \left[ \frac{1}{x^2}\delta^+\!\left(\frac{1}{x}\right)
+ \frac{1}{(1-x)^2}\delta^+\!\left(\frac{1}{1-x}\right) \right] +\mathcal{O}(\as^2).
\label{gamma-tmc-infinity}
\end{align}
The first boundary term is supported at $x=+\infty$ and the second at $x=-\infty$.  They do not contribute at finite
$x$, but they are part of the anomalous dimension when it is regarded as a distribution on the full line.

This distinction is directly related to quark-number conservation.  Using
\begin{equation}
\int_{-\infty}^{\infty}dx\, \frac{1}{x^2}\delta^+\!\left(\frac{1}{x}\right) = \int_{-\infty}^{\infty}dx\,
\frac{1}{(1-x)^2}\delta^+\!\left(\frac{1}{1-x}\right) =1,
\end{equation}
the coefficients in Eqs.~\eqref{gamma-tmc-finite} and \eqref{gamma-tmc-infinity} cancel exactly, so the zeroth moment
of the one-loop anomalous dimension vanishes,
\begin{equation}
\int_{-\infty}^{\infty}dx\, \Gamma_{\tilde q}^{\text{TMC}}(\as,x) =0+\mathcal{O}(\as^2).
\label{gamma-tmc-zero-moment}
\end{equation}
Thus the boundary distributions are invisible at finite $x$, but they are essential for the zeroth moment of the complete
anomalous dimension.  Equation~\eqref{gamma-tmc-zero-moment} shows that the quark-number condition imposed in defining
the TMC scheme is preserved by the RG evolution.

The RG equation for the renormalized quasi-PDF is therefore understood as an equation for distributions on the full line,
\begin{equation}
\frac{d}{d\ln\mu^2}\, \tilde q_{\text{TMC}}(x,p_z,\mu) = \Gamma_{\tilde q}^{\text{TMC}}(\as,x) \otimes
\tilde q_{\text{TMC}}(x,p_z,\mu).
\label{rg-qtmc}
\end{equation}

The matching relation in the TMC scheme is
\begin{equation}
\tilde q_{\text{TMC}}(x,p_z,\mu) = C_{\text{TMC}}(x,\mu,p_z)\otimes q_{\msbar}(x,\mu),
\label{match-rg-tmc}
\end{equation}
while the lightcone PDF satisfies the standard DGLAP evolution
equation~\cite{Gribov:1972ri,Gribov:1972rt,Dokshitzer:1977sg,Altarelli:1977zs}
\begin{equation}
\frac{d}{d\ln\mu^2}\,q_{\msbar}(x,\mu) = \Gamma_q(\as,x)\otimes q_{\msbar}(x,\mu), \qquad \Gamma_q(\as,x)
= \frac{\as C_F}{2\pi}P_{qq}(x) +\mathcal{O}(\as^2).
\label{rg-pdf}
\end{equation}
The matching coefficient then obeys the RG equation of Eq.~\eqref{eq:matching-rg-general}, specialized to
$S=\text{TMC}$,
\begin{equation}
\frac{d}{d\ln\mu^2} C_{\text{TMC}} = \Gamma_{\tilde q}^{\text{TMC}}\otimes C_{\text{TMC}}
- C_{\text{TMC}}\otimes \Gamma_q .
\label{rg-Ctmc-general}
\end{equation}
For the matching coefficient at finite $x$, the boundary-supported part of
$\Gamma_{\tilde q}^{\text{TMC}}$ does not enter the convolution with the
lightcone PDF.  Therefore the evolution of the one-loop matching
coefficient at finite $x$ is governed by
\begin{equation}
\frac{d}{d\ln\mu^2}C_{\text{TMC}}^{(1)}(x,\mu,p_z) = \frac{\as C_F}{2\pi} \left[ \frac{3}{2}\delta(1-x)-P_{qq}(x) \right],
\label{rg-Ctmc-one-loop}
\end{equation}
whereas the complete anomalous dimension of the quasi-PDF contains in addition the boundary term in
Eq.~\eqref{gamma-tmc-infinity}.  Thus the boundary terms at $x=\pm\infty$ do not modify the matching equation at finite
$x$, but they are required when the anomalous dimension is treated as a distribution on the full line.

This distinction is important conceptually.  The transverse cutoff $\Lambda$ regulates the UV loop momentum and 
produces explicit powers and logarithms of $\Lambda$.  The poles in Eq.~\eqref{gamma-tmc-infinity}, on the other 
hand, come from the  treatment of the distributions for large-$|x|$ boundary and are regularized by dimensional 
regularization.  They should not be confused with additional transverse-momentum divergences at finite $x$.   

More generally, the ordering of the convolutions in Eq.~\eqref{eq:matching-rg-general} reflects whether the
anomalous dimension acts after or before the matching map.
$\Gamma_{\tilde q}^{S}$ acts on the quasi-PDF, while $\Gamma_q$ acts on the lightcone PDF. Changing the 
renormalization prescription for the quasi-PDF changes $\Gamma_{\tilde q}^{S}$ and the matching coefficient 
accordingly.  The quasi-PDF, the matching coefficient, and the split between counterterm and remnant are therefore not 
physical separately.  Their scheme dependence cancels only in the factorization relation.

Equivalently, the inverse matching relation
\begin{equation}
q_{\msbar}(x,\mu) = C_{\text{TMC}}^{-1}(x,\mu,p_z) \otimes \tilde q_{\text{TMC}}(x,p_z,\mu)
\end{equation}
shows that the matched combination $C_{\text{TMC}}^{-1}\otimes \tilde q_{\text{TMC}}$ obeys the 
same DGLAP equation as the $\msbar$ PDF.  Conversely, $C_{\text{TMC}}\otimes q_{\msbar}$ evolves as the 
quasi-PDF in the TMC scheme.  The part of the anomalous dimensions of the quasi-PDF involving boundaries is 
invisible in the matching of the PDF at finite $x$, but it is required in the RG equation for the quasi-PDF as a distribution on
the full line. 

\section{Conclusion} \label{conc}

In this paper we studied the one-loop nonsinglet quark quasi-PDF in the transverse-momentum-cutoff scheme.  The 
purpose was to clarify how the UV contribution associated with the cutoff-regulated spatial operator is separated
from the renormalized quasi-PDF which enters the matching to the lightcone PDF.  The TMC scheme is useful 
for this question because the cutoff dependence appears explicitly in the one-loop calculation.  It therefore gives a simple 
setting in which the separation between the counterterm and the matching remnant can be followed directly in 
momentum-fraction space.

The separation has to be made more carefully than the statement that the counterterm contains the $\Lambda$-dependent 
terms alone. At finite $x$, the minimal TMC counterterm is indeed given by the explicit dependence on the
transverse-momentum cutoff.  At one loop this consists of the linearly divergent tadpole contribution and the logarithmic 
cutoff dependence from wave-function renormalization.  The tadpole contribution is represented by a double plus 
distribution centered at $x=1$, while the wave-function term gives the logarithmic $\mu$-dependence.

However, the quasi-PDF is a distribution on the whole real line.  After the explicit $\Lambda$-dependent terms are 
separated, the remaining one-loop contribution still has to be interpreted as a distribution on the whole real line.  
There are $1/x$-type tails in the regions $x>1$ and $x<0$, and expanding with respect to $\eps$ generates UV boundary 
poles at $x=\pm\infty$.  These poles are not additional divergences due to the transverse momentum cutoff at finite $x$, 
but boundary UV terms of the momentum-fraction distribution on the full line. They are therefore assigned to the TMC
counterterm together with the explicit cutoff-dependent terms.  Thus the one-loop structure is naturally organized in two steps — 
separating the explicit cutoff-dependent part $\tilde q_{\text{TMC,ct}}^{(1),\Lambda}$, which leaves an intermediate remnant not 
yet to be called the renormalized quasi-PDF, and then, once that remnant is written as a distribution on the whole line, absorbing the 
boundary poles at $x=\pm\infty$ into $\tilde q_{\text{TMC,ct}}^{(1),\infty}$. Only after both subtractions does one obtain the
TMC-renormalized quasi-PDF. This organization distinguishes the explicit UV cutoff contribution from the UV boundary contribution 
of the distribution on the full line.

The renormalized quasi-PDF obtained in this way retains the full collinear infrared pole in the physical region $0<x<1$.  
This pole is the same as the one in the lightcone PDF and cancels in the matching coefficient.  The boundary delta functions 
at $x=\pm\infty$, on the other hand, do not contribute to the matching convolution with the lightcone PDF at finite $x$.  
They have already been absorbed into the TMC counterterm. What remains in the matching coefficient is the finite 
full-line distribution, including the finite plus-distribution tails at infinity.

The same point is reflected in the RG behavior.  The anomalous dimension of the renormalized quasi-PDF is not only 
the contribution at finite $x$ generated by the logarithmic wave-function counterterm.  It also contains the boundary 
terms associated with the UV poles at $x=\pm\infty$. These terms are invisible in the matching coefficient at finite $x$, 
but they are part of the anomalous dimension of the quasi-PDF as a distribution on the whole real line. The matching 
coefficient compensates the difference between the quasi-PDF anomalous dimension in the TMC scheme and the 
anomalous dimension of the $\msbar$ lightcone PDF, so that the matched combination obeys the standard DGLAP evolution.
The finite nonlocal part of the result agrees with the standard one-loop full-line quasi-PDF matching 
result~\cite{Izubuchi:2018srq} once the explicit TMC-dependent contribution and the boundary UV terms are treated 
consistently.  This agreement is a check of the separation
prescription, not an identification of the TMC and $\msbar$ renormalization schemes.  The finite coefficient of $\delta(1-x)$ 
remains sensitive to the finite part assigned to the wave-function contribution. 

Although the TMC scheme is not used as the most practical prescription for lattice calculations, it provides a useful 
benchmark for understanding the renormalization of spatial correlation functions.  It makes explicit that the counterterm 
of a cutoff-regulated quasi-PDF has to be defined as a distribution on the full line. The explicit cutoff dependence removes 
the UV contribution of the transverse loop momentum, while the boundary terms at $x=\pm\infty$ complete the 
distributions of the UV divergence.  This separation clarifies how the cutoff-dependent contribution can be removed 
without disturbing the collinear infrared structure required for LaMET matching.

It would be interesting to apply the idea of scheme separation to other renormalization schemes, so that the part of
the renormalized quasi-PDF and the matching coefficient tied to the particular choice of scheme can be cleanly
identified and separated from the physical content relevant for matching. In contrast to the $\msbar$
scheme and the TMC scheme,  other renormalization schemes may have the renormalization prescriptions which are
hard to disentangle. We will consider other renormalization schemes such as the RI/MOM scheme and ask the question on
the scheme separation in the renormalized quasi-PDF and the matching coefficient in the near future.
 
 \appendix

\section{Distributions on the full line with endpoint and infinity}
\label{app:full-line-distribution}

We summarize the conventions for the distributions used in the expression for
the renormalized quasi-PDF and the matching coefficient on the full line.
Two different endpoint structures appear.  The first is the ordinary endpoint
at $x=1$, where plus and double plus distributions are defined.  The second
is the behavior at $x=\pm\infty$, which is specific to the quasi-PDF on the
full real line and produces boundary distributions when the large-$|x|$
tails are treated before taking the limit $\epsilon\to0$.  We then extract
the endpoint-local $\delta(1-x)$ coefficient needed in the finite remnant.

\subsection{Endpoint distributions at $x=1$}
\label{app:endpoint-distributions}

Here we consider the distributions at the endpoint $x=1$.
For an interval $D$ containing $x=1$, we define the plus distribution as
\begin{equation}
\int_D dx\, [f(x)]_{+(1)}^D h(x) \equiv \int_D dx\, f(x)\,[h(x)-h(1)] ,
\label{eq:app-plus-def}
\end{equation}
where $h(x)$ is a smooth test function.  For the tadpole contribution we also encounter a second-order endpoint singularity.  
We define the double plus distribution as
\begin{equation}
\int_D dx\, \left[\frac{1}{(1-x)^2}\right]_{++(1)}^D h(x) \equiv \int_D dx\,\frac{h(x)-h(1)-(x-1)h'(1)}{(1-x)^2}.
\label{eq:app-double-plus-def}
\end{equation}
With this convention the linear tadpole counterterm is written entirely as the double plus distribution centered at $x=1$.  
This definition fixes the local endpoint part of the second-order plus
distribution.  To see this, consider another convention differing from
Eq.~\eqref{eq:app-double-plus-def} by local terms at $x=1$,
\begin{equation}
  \left[\frac{1}{(1-x)^2}\right]_{++(1)}^{\rm alt}  =   \left[\frac{1}{(1-x)^2}\right]_{++(1)}^{\rm here}
  + A\,\delta(1-x)+B\,\delta'(1-x).
\end{equation}
Its action on a test function differs by 
\begin{equation}
  A\,\phi(1)-B\,\phi'(1).
\end{equation}
Thus a change of convention only reshuffles local endpoint contributions between the definition of the double plus 
distribution and explicitly written $\delta(1-x)$, $\delta'(1-x)$ terms.  In the convention used in this paper we choose 
the distributions defined in Eq.~\eqref{eq:app-double-plus-def}, hence the linearly divergent tadpole counterterm is 
written entirely as a double plus distribution without additional explicit local endpoint terms.

\subsection{Large-$|x|$ tails and boundary distributions}
\label{app:infinity-distribution-form}

We next discuss the regions $x>1$ and $x<0$.  Before the large-$|x|$ tails are rearranged, the relevant functions are
\begin{align}
f_>(x) &= \frac{1+x^2}{1-x}\ln\frac{x}{x-1}+1, \qquad x>1, \nonumber\\
f_<(x) &= \frac{1+x^2}{1-x}\ln\frac{1-x}{-x}-1, \qquad x<0.
\label{eq:app-fpm-def}
\end{align}
These are  the functions appearing in Eq.~\eqref{qrem-finite-x} for $x>1$ and $x<0$ respectively. Their large-$|x|$ 
behavior is
\begin{align}
f_>(x) &= -\frac{3}{2x}+\mathcal{O}(x^{-2}), \qquad x\to+\infty, \nonumber\\
f_<(x) &= -\frac{3}{2(1-x)}+\mathcal{O}(x^{-2}), \qquad x\to-\infty.
\label{eq:app-fpm-asym}
\end{align}
The second form is convenient because $1/(1-x)\to0^+$ as $x\to-\infty$.
In the main text we added and subtracted the asymptotic forms such that the terms inside the brackets fall as $1/x^2$ at infinity.
\begin{align}
R_{\text{rem}}^{\text{out}}(x) =& \left[ f_>(x)+\frac{3}{2x} \right]_{+(1)}^{[1,\infty]} -\frac{3}{2x}\,\theta(x-1) \nonumber\\
&+ \left[ f_<(x)+\frac{3}{2(1-x)} \right]_{+(1)}^{[-\infty,0]} -\frac{3}{2(1-x)}\,\theta(-x).
\label{eq:app-Rout-addsubtract}
\end{align}
This representation is useful for the finite remnant and for the matching coefficient.  However, if the quasi-PDF is regarded 
as a distribution on the full line before the UV subtraction is made, the two explicit tails in 
Eq.~\eqref{eq:app-Rout-addsubtract} also contain boundary UV poles at $x=\pm\infty$.

The boundary delta functions are defined by their action on a test function $h(x)$ as
\begin{align}
\int_1^\infty dx\, \frac{1}{x^2}\delta^+\!\left(\frac1x\right)h(x) &=h(+\infty), \nonumber\\
\int_{-\infty}^0 dx\, \frac{1}{(1-x)^2} \delta^+\!\left(\frac{1}{1-x}\right)h(x)
&=h(-\infty).
\label{eq:app-delta-infinity-def}
\end{align}
The superscript $+$ means that the argument of the delta function approaches zero from the positive side. The 
corresponding plus distributions at infinity are defined by
\begin{align}
\int_1^\infty dx\, \left(\frac1x\right)_{+(\infty)}^{[1,\infty]}h(x) &= \int_1^\infty dx\,\frac{h(x)-h(+\infty)}{x},
\nonumber\\
\int_{-\infty}^0 dx\, \left(\frac{1}{1-x}\right)_{+(-\infty)}^{[-\infty,0]}h(x) &= \int_{-\infty}^0 dx\,
\frac{h(x)-h(-\infty)}{1-x}.
\label{eq:app-plus-infinity-def}
\end{align}
Keeping the dimensional regulator until the boundary expansion is performed, as in the full-line treatment of 
Ref.~\cite{Izubuchi:2018srq}, one obtains
\begin{equation}
\theta(x-1)x^{-1-2\epsilon} = \frac{1}{2\euv} \frac{1}{x^2}\delta^+\!\left(\frac1x\right) +
\left(\frac1x\right)_{+(\infty)}^{[1,\infty]} +\mathcal{O}(\epsilon),
\label{eq:app-plus-infinity-right}
\end{equation}
and
\begin{equation}
\theta(-x)(1-x)^{-1-2\eps} = \frac{1}{2\euv} \frac{1}{(1-x)^2}\delta^+\!\left(\frac{1}{1-x}\right) +
\left(\frac{1}{1-x}\right)_{+(-\infty)}^{[-\infty,0]} +\mathcal{O}(\eps).
\label{eq:app-plus-infinity-left}
\end{equation}
The factor $1/2$ in the pole comes from the power $1+2\eps$.  To determine the finite boundary terms, however, the
coefficient of the large-$|x|$ tail must also be retained through first order in $\eps$.  From the regulated expression in
Eq.~\eqref{qrem-unexpanded}, the three terms in the square brackets have, for $x\to+\infty$, the coefficients
\begin{equation}
-\frac{1}{1-2\eps}, \qquad \frac{1-\eps}{2}, \qquad -1,
\end{equation}
respectively, multiplying $x^{-1-2\eps}$.  Their sum is
\begin{equation}
-\frac{3}{2}-\frac{5}{2}\eps+\mathcal{O}(\eps^2).
\end{equation}
The common prefactor in Eq.~\eqref{qrem-unexpanded} has the expansion
\begin{equation}
\left(\frac{\mu^2 e^{\gamma_{\text E}}}{p_z^2}\right)^\eps
\frac{\Gamma(1/2+\eps)}{\sqrt{\pi}}
=1+\eps\ln\frac{\mu^2}{4p_z^2}+\mathcal{O}(\eps^2).
\end{equation}
Therefore the regulated large-$x$ tail is
\begin{equation}
\left[-\frac{3}{2}-\eps A+\mathcal{O}(\eps^2)\right]\theta(x-1)x^{-1-2\eps}, \qquad
A\equiv \frac{3}{2}\ln\frac{\mu^2}{4p_z^2}+\frac{5}{2}.
\label{eq:app-regulated-tail-right}
\end{equation}
The same coefficient is obtained for $x\to-\infty$, with $x^{-1-2\eps}$ replaced by $(1-x)^{-1-2\eps}$,
\begin{equation}
\left[-\frac{3}{2}-\eps A+\mathcal{O}(\eps^2)\right]\theta(-x)(1-x)^{-1-2\eps}.
\label{eq:app-regulated-tail-left}
\end{equation}
The $\mathcal{O}(\eps)$ term in the coefficient cannot be dropped before the boundary expansion because it multiplies the
$1/\euv$ pole in Eqs.~\eqref{eq:app-plus-infinity-right} and \eqref{eq:app-plus-infinity-left}.  We thus obtain
\begin{align}
\left(-\frac{3}{2}-\eps A\right)\theta(x-1)x^{-1-2\eps} 
&= -\frac{3}{4\euv}\frac{1}{x^2}\delta^+\!\left(\frac1x\right)
-\frac{3}{2}\left(\frac1x\right)_{+(\infty)}^{[1,\infty]}
-\frac{A}{2}\frac{1}{x^2}\delta^+\!\left(\frac1x\right)
+\mathcal{O}(\eps), \nonumber\\
\left(-\frac{3}{2}-\eps A\right)\theta(-x)(1-x)^{-1-2\eps}
&= -\frac{3}{4\euv}\frac{1}{(1-x)^2}\delta^+\!\left(\frac{1}{1-x}\right)
-\frac{3}{2}\left(\frac{1}{1-x}\right)_{+(-\infty)}^{[-\infty,0]} \nonumber \\
&\qquad -\frac{A}{2}\frac{1}{(1-x)^2}\delta^+\!\left(\frac{1}{1-x}\right)
+\mathcal{O}(\eps).
\label{eq:app-tail-boundary-expansion}
\end{align}
Thus the boundary UV pole from the two semi-infinite regions is
\begin{equation}
 -\frac{3}{4\euv} \left[ \frac{1}{x^2}\delta^+\!\left(\frac1x\right) + \frac{1}{(1-x)^2} \delta^+\!\left(\frac{1}{1-x}\right)
\right].
\label{eq:app-Ruv-infinity}
\end{equation}
This is the origin of the boundary part of the TMC counterterm and of $\Gamma_{\tilde q}^{\mathrm{TMC,\infty}}$ in the
RG equation.  After this UV pole is assigned to the counterterm, the finite large-$|x|$ contribution contains, in addition
to the plus distributions at infinity, the boundary term
\begin{equation}
-\frac{1}{2}\left(\frac{3}{2}\ln\frac{\mu^2}{4p_z^2}+\frac{5}{2}\right)\left[
\frac{1}{x^2}\delta^+\!\left(\frac1x\right)
+\frac{1}{(1-x)^2}\delta^+\!\left(\frac{1}{1-x}\right)
\right].
\label{eq:app-finite-boundary-term}
\end{equation}

\subsection{Extraction of the coefficient for $\delta(1-x)$}
\label{app:endpoint-coefficient}

The finite function appearing in the remnant for $0<x<1$ in Eq.~\eqref{finalqrem} is
\begin{equation}
F(x) = \frac{1+x^2}{1-x} \ln\frac{4x(1-x)p_z^2}{\mu^2} - \frac{x(1+x)}{1-x}, \qquad 0<x<1 .
\label{eq:app-F-physical}
\end{equation}
The plus distribution $[F(x)]_{+(1)}^{[0,1]}$ is defined by Eq.~\eqref{eq:app-plus-def}. Equivalently, one may introduce 
an endpoint cutoff $0<x<1-\rho$, with $\rho\to0^+$, and isolate the finite term multiplying $h(1)$.

We write
\begin{equation}
\ln\frac{4x(1-x)p_z^2}{\mu^2} = \ln\frac{4p_z^2}{\mu^2} +\ln x+\ln(1-x).
\label{eq:app-Lz-def}
\end{equation}
The term proportional to $\ln x$ is regular at $x=1$, since
\begin{equation}
\lim_{x\to1}\frac{\ln x}{1-x}=-1.
\end{equation}
The term proportional to $\ln(1-x)$ is singular at the endpoint.  In the convention used here this logarithmic singularity 
is kept inside the plus distribution and is not rewritten as an additional explicit $\delta(1-x)$ term.  The explicit finite 
coefficient of $\delta(1-x)$ is therefore obtained from the finite endpoint parts of the $\ln(4p_z^2/\mu^2)$ term and the 
rational term.

The coefficient of the $\ln(4p_z^2/\mu^2)$ term follows from
\begin{equation}
I_1(\rho) = \int_0^{1-\rho}dx\,\frac{1+x^2}{1-x} = -2\ln\rho-\frac32+\mathcal{O}(\rho).
\label{eq:app-I1}
\end{equation}
Thus its finite endpoint contribution is $-\frac{3}{2}\ln(4p_z^2/\mu^2)$, which we write as 
$\frac{3}{2}\ln(\mu^2/4p_z^2)$ in the final result. For the rational term,
\begin{equation}
I_2(\rho) = \int_0^{1-\rho}dx\, \left[-\frac{x(1+x)}{1-x}\right] = 2\ln\rho+\frac52+\mathcal{O}(\rho).
\label{eq:app-I2}
\end{equation}
The logarithmic divergence is part of the endpoint subtraction, while the finite rational contribution is $5/2$.  Hence, 
before including the finite part of the wave-function remnant,
\begin{equation}
C_\delta^{(0<x<1)} = \frac32\ln\frac{\mu^2}{4p_z^2} +\frac{5}{2}.
\label{eq:app-Cdelta-before-wf}
\end{equation}
The contribution of the wave-function renormalization to the renormalized quasi-PDF  is
\begin{equation}
\tilde q_{\text{wf}}^{(1),\text{ren}}(x,\mu) = \asc\, \frac{1}{2\eir}\delta(1-x).
\label{eq:app-wf-rem}
\end{equation}
The finite wave-function term is assigned to the counterterm by the quark-number condition
in Eq.~\eqref{tmc-quark-number}.  It therefore does not shift the finite endpoint coefficient, and
\begin{equation}
C_\delta^{\text{ren}} = \frac32\ln\frac{\mu^2}{4p_z^2} +\frac{5}{2}.
\label{eq:app-Cdelta-final}
\end{equation}

The collinear pole also contains a local contribution through the quark
splitting function.  With the plus-distribution convention of
Eq.~\eqref{eq:app-plus-def}, we use
\begin{equation}
P_{qq}(x) = \left[\frac{1+x^2}{1-x}\right]_{+(1)}^{[0,1]} = 2\left[\frac{1}{1-x}\right]_{+(1)}^{[0,1]} -(1+x)
+\frac{3}{2} \delta(1-x).
\label{eq:app-Pqq-def}
\end{equation}
Therefore the pole term $-P_{qq}(x)\theta(x)\theta(1-x)/\eir$ contains
\begin{equation}
-\frac{3}{2\eir}\delta(1-x).
\end{equation}
Combining the pole term and the finite term, the part proportional to $\delta(1-x)$ in the remnant is
\begin{equation}
\frac{\as C_F}{2\pi} \left[ -\frac{3}{2\eir} + \frac32\ln\frac{\mu^2}{4p_z^2} +\frac{5}{2} \right]\delta(1-x).
\label{eq:app-endpoint-local-final}
\end{equation}
This equation only displays the local part of Eq.~\eqref{finalqrem}. It is not
an additional contribution.
  
%


\end{document}